\documentstyle[12pt,aaspp4]{article}
\makeatletter

\@ifundefined{chapter}{\def\thebibliography#1{\section*{References\@mkboth
  {REFERENCES}{REFERENCES}}\list
  {\relax}{\setlength{\labelsep}{0em}
        \setlength{\itemindent}{-\bibhang}
        \setlength{\itemsep}{0pt}
        \setlength{\parsep}{0pt}
        \setlength{\leftmargin}{\bibhang}}
    \def\newblock{\hskip .11em plus .33em minus .07em}
    \sloppy\clubpenalty4000\widowpenalty4000
    \sfcode`\.=1000\relax}}%
{\def\thebibliography#1{\chapter*{Bibliography\@mkboth
  {BIBLIOGRAPHY}{BIBLIOGRAPHY}}\list
  {\relax}{\setlength{\labelsep}{0em}
        \setlength{\itemindent}{-\bibhang}
        \setlength{\itemsep}{0pt}
        \setlength{\parsep}{0pt}
        \setlength{\leftmargin}{\bibhang}}
    \def\newblock{\hskip .11em plus .33em minus .07em}
    \sloppy\clubpenalty4000\widowpenalty4000
    \sfcode`\.=1000\relax}}

\newlength{\bibhang}
\let\@internalcite\cite
\def\cite{\let\@citeleft(\let\@citeright)%
    \@ifstar{\citeyear}{\citefull}}
\def\acite{\let\@citeleft\relax\let\@citeright\relax%
    \@ifstar{\citeyear}{\acitefull}}
\def\citenp{\let\@citeleft\relax\let\@citeright\relax
    \@ifstar{\citeyear}{\citefull}}
\def\citefull{\def\astroncite##1##2{##1~##2}\@internalcite}
\def\citeyear{\def\astroncite##1##2{##2}\@internalcite}
\def\acitefull{\def\astroncite##1##2{##1~(##2)}\@internalcite}

\def\@citex[#1]#2{\if@filesw\immediate\write\@auxout{\string\citation{#2}}\fi
  \def\@citea{}\@cite{\@for\@citeb:=#2\do
    {\@citea\def\@citea{; }\@ifundefined
       {b@\@citeb}{{\bf ?}\@warning
       {Citation `\@citeb' on page \thepage \space undefined}}%
{\csname b@\@citeb\endcsname}}}{#1}}

\def\@cite#1#2{\@citeleft#1\if@tempswa , #2\fi\@citeright}
\def\@biblabel#1{}

\makeatother

\newcommand{\PSbox}[3]{\mbox{\rule{0in}{#3}\includegraphics{#1}\hspace{#2}}}
\newcommand{\FigNum}[1]{\unitlength 1pt \begin{picture}(55,10)(-400,35) 
                        \put(0,0){Figure #1}
                        \end{picture}}
\newcommand{\msun}{$M_\odot$} 
 
\newcommand{\persec}{\mbox{$\second^{-1}$}}
\newcommand{\percm}{\mbox{$\cm^{-2}$}}
\newcommand{\ppm}{\mbox{$\pm$}}
\newcommand{\cgsflux}{\erg\percm\persec}
\newcommand{\cgslum}{\erg\persec}

\newcommand\approxgt{\mbox{$^{>}\hspace{-0.24cm}_{\sim}$}}
\newcommand\approxlt{\mbox{$^{<}\hspace{-0.24cm}_{\sim}$}}
\def\etal{{et~al.}}

\newcommand{\nh}{\mbox{$N_{\rm H}$}}
\newcommand{\nhtt}{\mbox{$N_{\rm H, 22}$}}
\newcommand{\ud}[2]{\mbox{$^{+ #1}_{- #2}$}}
\newcommand{\ee}[1]{\mbox{$10^{#1}$}}
\newcommand{\tee}[1]{\mbox{$\times 10^{#1}$}}

\newcommand\lxlbol{$L_{X}$/$L_{\rm bol}$}

\newcommand{\perval}[2]{{#1\mbox{$^{#2}$}}} 

\def\chisqrnu{\mbox{$\chi^2_\nu$}}
\def\x1608{{4U~1608$-$522}}

\def\cenx4{{Cen~X$-$4}}
\def\aql{{Aql~X$-$1}}
\def\saxj1808{{SAX J1808.4$-$3658}}

\newcommand{\cm}{\mbox{$\rm\,cm$}}

\newcommand{\second}{\mbox{$\rm\,s$}}

\newcommand{\erg}{\mbox{$\rm\,erg$}}

\newcommand{\kpc}{\mbox{$\rm\,kpc$}}

\newcommand{\kteff}{\mbox{$kT_{\rm eff}$}}
\newcommand{\kteffinfty}{$kT_{\rm eff, \infty}$}

\newcommand{\rinfty}{\mbox{$R_{\infty}$}}

\newcommand{\chandra}{{\em Chandra\/}}

\newcommand{\rosat}{{\em ROSAT\/}}
\newcommand{\asca}{{\em ASCA\/}}
\newcommand{\rxte}{{\em RXTE\/}}

\newcommand{\beppo}{{\em BeppoSAX\/}}
\newcommand{\sax}{\beppo}

\def\fxpl{\mbox{$F_{X,pl}$}}
\def\object3{CXOU~132619.7-472910.8}
\def\rrr{R01}

\begin{document}

\title{Variable Thermal Emission from Aql X-1 in Quiescence}

\author{Robert E. Rutledge\altaffilmark{1}, 
Lars Bildsten\altaffilmark{2}, Edward F. Brown\altaffilmark{3}, 
George G. Pavlov\altaffilmark{4}, 
\\ and Vyacheslav  E. Zavlin\altaffilmark{5}}
\altaffiltext{1}{
Department of Physics, Mathetmatics and Astronomy, California
Institute of Technology, MS 130-33, Pasadena, CA 91125;
rutledge@tapir.caltech.edu}
\altaffiltext{2}{
Institute for Theoretical Physics and Department of Physics, Kohn Hall, University of 
California, Santa Barbara, CA 93106; bildsten@itp.ucsb.edu}
\altaffiltext{3}{
Enrico Fermi Institute, 
University of Chicago, 
5640 South Ellis Ave, Chicago, IL  60637; 
brown@flash.uchicago.edu}
\altaffiltext{4}{
The Pennsylvania State University, 525 Davey Lab, University Park, PA
16802; pavlov@astro.psu.edu}
\altaffiltext{5}{ 
Max-Planck-Institut f\"ur Extraterrestrische Physik, D-85748 Garching,
Germany; zavlin@xray.mpe.mpg.de}

\begin{abstract}

We obtained four \chandra/ACIS-S observations beginning two weeks
after the end of the November 2000 outburst of the neutron star (NS)
transient \aql . Over the five month span in quiescence, the X-ray
spectra are consistent with thermal emission from a NS with a pure
hydrogen photosphere and \rinfty=15.9\ud{0.8}{2.9} (d/5~kpc) km at the
optically implied X-ray column density. We also detect a hard
power-law tail during two of the four observations. The intensity of
\aql\ first decreased by 50\ppm4\% over three months, then increased
by 35\ppm5\% in one month, and then remained constant ($<$6\% change)
over the last month.  These variations in the first two observations
cannot be explained by a change in the power-law spectral component,
nor in the X-ray column density. Presuming that \rinfty\ is not
variable and a pure hydrogen atmosphere, the long-term changes can
only be explained by variations in the NS effective temperature, from
\kteffinfty=130\ud{3}{5} eV, down to 113\ud{3}{4} eV, finally
increasing to 118\ud{9}{4} eV for the final two observations. During
one of these observations, we observe two phenomena which were
previously suggested as indicators of quiescent accretion onto the NS:
short-timescale ($<$\ee{4} sec) variability (at 32\ud{8}{6}\% rms),
and a possible absorption feature near 0.5 keV. The possible
absorption feature can potentially be explained as due to a
time-variable response in the ACIS detector.  Even so, such a feature
has not been detected previously from a NS, and if confirmed and
identified, can be exploited for simultaneous measurements of the
photospheric redshift and NS radius.

\end{abstract}

\keywords{stars: atmospheres  --- stars: individual (Aql X-1) --- stars: neutron --- x-rays: binaries}

\section{Introduction}

Brown, Bildsten \& Rutledge \cite*[BBR98 hereafter]{brown98} showed
that the core of a transiently accreting neutron star (NS), such as
Aql X-1 (for reviews of transient neutron stars, see
\citenp{chen97,campana98b}), is heated to a steady-state temperature
by nuclear reactions deep in the crust during the accretion outbursts.
The timescale for the core to reach this steady-state is $\sim$\ee{4}
yr (see also \citenp{colpi00}), after which the NS emits a thermal
luminosity in quiescence of (BBR98)
\begin{equation}
\label{eq:brown}
L_q = 8.7\times10^{33} \left(\frac{\langle \dot{M} \rangle}{10^{-10}
M_\odot {\rm yr}^{-1}}\right) \frac{Q}{1.45 {\rm MeV}/m_p} \; \; {\rm ergs \; s}^{-1},
\end{equation}
where $\langle \dot{M} \rangle$ is the time-averaged mass-accretion
rate onto the NS, and $Q$ is the amount of heat deposited in the crust
per accreted nucleon (\citenp{haensel90}; see
\citenp{bildstenrutledge00} for a discussion).  Rutledge \etal\
\cite*[\rrr\ hereafter]{rutledge01b} showed that, in the case of \aql,
the quiescent X-ray luminosity was within observational uncertainties
of that predicted on the basis of its time-averaged accretion
luminosity. This observation, and similar observations of other
quiescent transient neutron stars (qNSs), strongly supports the
scenario that deep crustal heating provides a ``rock bottom''
quiescent luminosity.

 This quiescent thermal emission does not preclude that accretion
onto the NS surface (at a rate $\dot M_c$) continues in quiescence at
a substantially lower rate than during outburst. \acite{narayan97a}
argued that the analogous black hole (BH) systems accrete during
quiescence via an advective flow that radiates inefficiently and makes
quiescent BHs dim or non-detectable.  This same accretion flow onto
the NSs should be observable, with efficiencies $\epsilon=0.20$
($L_{\rm X}\sim \epsilon \dot{M}_c c^2$). Statistical comparisons show
that the BHs are, on average, less luminous than NSs by a factor of
$\sim$10 \cite{barret94,narayan97a,asai98}, although in individual
cases they may be more luminous (V404 Cyg is an example of a BH which
is more luminous than most qNSs; \citenp{kong01}).
While this average difference in luminosity argues for a difference
between the emission properties of NSs and of BHs, perhaps due to the
presence of a surface in the former, detailed X-ray spectroscopy is
still required to distinguish quiescent NSs from quiescent BHs in
individual cases where the compact object mass is not measured nor
type-I X-ray bursts observed
\cite{rutledge00}.

Assessing the relevance of accretion as a source of the NS quiescent
emission is hampered by the difficulty of predicting $\dot{M}_{c}$.
In the context of an advection-dominated accretion flow (ADAF;
\citenp{narayan94}), $\dot{M}_c$ is set by the evaporation rate from
the disk which is in turn related to the unknown disk truncation
radius.  Similar limitations apply for other low $\dot{M}$ solutions,
such as the Adiabatic Inflow/Outflow Solution (ADIOS;
\citenp{blandford99}). We are thus searching for direct indicators of
quiescent accretion onto the NS: short term (timescales much less than
the orbital period) intensity variability, and evidence of metals in
the photosphere (BBR98; \citenp{bildstenrutledge00}).

An intensity decrease of 40\ppm8\% across 5 yrs in Cen X-4 may well be
explained as crustal cooling from a long ($\sim$100 yr) recurrence
time NS \cite{rutledge01}.  Previous observations have shown shorter
term variability (\approxgt 1 day). In Cen~X-4 \cite{campana97}, the
intensity decreased by a factor of $\sim3$ over 4 days\footnote{Our
re-analysis of the data found that the ROSAT/HRI countrate decreased
by a factor of 4.1\ppm0.6 over 4.4 days.}  in 1997, approximately 18
years after the previous known outburst.  Since, in crustal cooling
variability, the timescale for variability scales with the time since
the most recent outburst, four-day variability is much too short a
timescale to occur 18 years post-outburst to be explained by this
mechanism \cite{ushomirsky01}.

A second means by which active accretion onto the NS would be
indicated during quiescence is the presence of metals in the
photosphere. For very low accretion rates (an order of magnitude
estimate is $\dot{M}_c\ll$\ee{-13} \msun \perval{yr}{-1}) a transient
NS will have a pure H atmosphere because the heavy element settling
time is $\sim 10$~s at the photosphere \cite{bildsten92}.   At
accretion rates above this, metals continuously populate the photosphere
at such densities that their presence may be observable as absorption
lines or edges (see Fig.~\ref{fig:slava}). 

Aql X-1 has been detected in X-ray quiescence six times: once with the
\rosat/HRI and twice with \rosat/PSPC \cite{verbunt94}, once with
\asca\ \cite{asai98}, once with \sax\ \cite{campana98a}, and once with
\chandra\ (\rrr).  In all but the last (\rrr), the emission was
interpreted as blackbody, although a significant power-law component
was detected by \acite{campana98a}, which dominated the emission above
$\sim$2 keV, similar to a component observed from Cen~X-4
\cite{asai96a,campana00,rutledge01}.  \rrr\ examined a number of
spectral models, discarding them in favor of a H atmosphere spectrum
associated with emission from the NS photosphere.

We report here three additional \chandra\ detections in quiescence.
In three of four \chandra\ observations of \aql\ in quiescence, we
detect neither short term variability, nor absorption lines; however,
in the fourth observation, we detect both variability and a possible
absorption line.  While other explanations are not excluded, the
presence of both short-term (\approxlt \ee{4} sec) variability and
absorption during the fourth observation, combined with the absence of
both during the other three observations, is evidence for active
accretion onto the NS in quiescence.  The modest significance
($5\sigma$) of the absorption line does not permit detailed
spectroscopic study; however, its presence--- should it be confirmed
with higher S/N observations---opens up the opportunity to measure the
redshift at the NS photosphere.  Simultaneous measurements of the
redshift, which is a function of $M/R$, and $R$ would constrain the NS
equation of state.

We begin, in \S~\ref{sec:aql-x-1}, by reviewing the distance and
reddening to \aql.  We then describe, in \S~\ref{sec:obs}, our
observations,  timing analysis, and spectral analysis.
Section~\ref{sec:con} discusses our interpretation of the spectra and
its implications.

\section{Aql X-1: Distance and Reddening}\label{sec:aql-x-1}

Callanan, Filippenko \& Garcia \cite*{callanan99} showed that the
optical counterpart to Aql X-1 is a faint star 0.46\arcsec\ from the
previously mis-identified counterpart. This led to the counterpart's
identification as a late type star (spectral type K7 to M0) with a
quiescent magnitude $V=21.6$ with reddening of $E(B-V)=0.5$\ppm0.1
\cite{chev99}.  For standard conversions\footnote{The formal uncertainty
in this conversion is 2\%, however there are clear systematic deviations
of 25\% to factors of $\sim$few in some X-ray sources; this may be due
in part to the assumed X-ray spectral model.} ($A_V/E(B-V)$=3.1, 
\citenp{fitzpatrick99}; \nh/$A_V$=1.79\tee{21} \perval{cm}{-2}
\perval{mag}{-1}, \citenp{predehl95}), this corresponds to an X-ray
column density of \nhtt=0.28\ppm0.06 (\nh=\nhtt \ee{22}
\perval{cm}{-2}).  \acite{thorstensen78} note that a nearby (1.4\arcmin)
B-type star has an optical reddening $E(B-V)$=0.73~mag (no uncertainty
given), implying \nhtt=0.40.  These values are comparable to the
integrated galactic HI measurements in the direction of \aql\ of
\nhtt=0.34\ppm0.01 (\citenp{dickey90}, from
W3nH\footnote{http://heasarc.gsfc.nasa.gov}; the uncertainty was found
as $\sigma_{1^\circ}/\sqrt{N}$, where $\sigma_{1^\circ}$ is the
dispersion of the $N$ measurements within 1 degree, and $N=7$).  These
\nh\ values are listed in Table~\ref{tab:nh}.  These values will be
used for comparison with the results from spectral fitting in the
following section.  

The orbital period has been measured at $P_{\rm orb}=18.95 \ {\rm hr}$
via photometric observations both in outburst
\cite{chevalier98,garcia99b} and quiescence \cite{welsh00}.  While
Chevalier et al. \cite*{chev99} estimated the distance to the binary
as 2.5 kpc, a re-examination of the arguments producing this distance
(which neglected the need for Roche-lobe overflow in the primary)
finds that it is between 4 and 6.5 kpc (\rrr); we adopt 5 kpc as our
fiducial distance.  This 20\% uncertainty is neglected in our quoted
uncertainties in \rinfty, which only take into account the statistical
uncertainty in our spectral fits. The system's orbital inclination is
estimated to be $>36^\circ$ \cite{welsh00}.

\section{Observations and Analysis}
\label{sec:obs}

\aql\ was observed on four occasions, with identical instrumental
set-up using the \chandra/ACIS-S detector, backside-illuminated
chip. The X-ray source was placed 4\arcmin\ off-axis and the read-out
limited to 1/8 of the chip area with time resolution of 0.44104
seconds; these settings mitigate pile-up to $<1\%$ of the detected
photons.  In all observations, source counts were extracted from an 8
pixel radius about the source position; background was taken from an
annulus of inner and outer radius of 10 and 50 pixels,
respectively. The spacecraft roll angle differed between each
observation.  The X-ray source, however, fell on the same physical
pixels which, due to the spacecraft dither (used to average over QE
non-uniformity and bad pixels) and larger PSF, spanned $\sim$ 50
detector pixels in both directions across the chip. In each of the
four observations, the counts were centered near ({\tt chipx,
chipy})=(690,710), with deviations between observations of \ppm4
pixel; there are no known bad pixels within 50 pixels of this location.
The expected background countrate in the source region was always
$<$1\%.  Analysis of the first of these observations is described
elsewhere (\rrr).  An observation list is in Table~\ref{tab:obs}.

For the analysis we use CIAO~2.2.0.1 with CALDB~2.9; the ACIS-S3 chip
calibration was improved in CALDB
2.7\footnote{http://cxc.harvard.edu/cal/Links/Acis/acis/Cal\_prods/matrix/matrix.html};
analyses of astrophysical sources shows no evidence of systematic
deviations $>$15\%.  However, the effects of a recently reported time
dependence in the ACIS response is discussed in Sec.~\ref{sec:resp}.

The first observation triggered as a Target of Opportunity, following
the end of a bright X-ray outburst (see Fig.~\ref{fig:lc}) -- the
brightest observed yet with RXTE/ASM \cite{jain00,rutledge00aqlatel}.
The other three observations followed after a space of 2, 3 and 4
months from the first observation.  The countrates in
Table~\ref{tab:obs}, in three separate passbands, show significant
variability between observations 1, 2 and 3; observations 3 and 4 show
identical countrates in all passbands.  In the 0.5-10 keV passband,
the countrate decreased by 51\ppm8\% between observations 1 and 2
(over a period of 81 days), increased by 35\ppm4\% between obs. 2 and
3 (across 32 days), and then remained consistent with constant
(\ppm6\%) across 28 days).

\subsection{Short-Term ($<$\ee{4} sec) Variability}

Using the same approach as in \rrr, we produced power density spectra
(PDS) using all counts detected (no energy constraints, corresponding
roughly to 0.2-10 keV) to search for broad-band variability (see
\citenp{rogg} for the use of PDS in measuring broad-band variability in
X-ray sources).  We produced Fourier transforms of the time-series
data, with time resolution of 0.44104 sec \cite{press}.  From these,
we produced the PDS (the sum of the squares of the Fourier components
as a function of frequency), and rebinned them logarithmically in
frequency.  The resulting PDS are shown in Fig.~\ref{fig:pds}. 

Visual examination of the PDS shows no evidence for broad-band power
in observations 1-3 (there is marginally significant variability in
observation 3, near 2\tee{-4} Hz); however, there is clear excess
power at low frequencies during observation 4.  We modeled the data as
a power-law in frequency, plus a constant (Poisson) level ($P(\nu)=A
\nu^{-\alpha} + C$); the best fit model gives a slope
$\alpha=1.8\ppm0.4$ and an root-mean-square (RMS) variability of
32\ud{8}{6}\%, integrated between 0.0001-1 Hz (0.2-10 keV).  Similar
modeling to observations 1-3 (holding the power-law slope fixed at the
best-fit value of observation 4) provide 3$\sigma$ upper-limits
between 18-29\% rms during these observations (see
Table~\ref{tab:var}). For comparison, we show the lightcurve of obs. 4
in Fig.~\ref{fig:obs4lc}.  This is the first observation of short
timescale ($<10^4\,\mathrm{s}$) variability observed from the
intrinsic spectrum of a quiescent NS transient (that is, not
attributable to changes in the intervening X-ray column density).

We estimate (based on spectral fiting, below) that the power-law
component only contributes 12\% of the counts in the detector in the
0.2-10 keV energy range.  This is insufficient to produce 32\%
variability.  We conclude that the variability is in the thermal
component.

Since the ratio of X-ray to bolometric luminosity
log(\lxlbol)=0.5\ppm0.2 from \aql\ in quiescence is more than three
orders of magnitude greater than observed from coronal emission from
the analogous RS CVn systems \cite{bildstenrutledge00}, we exclude the
possibility that observed variability (at $\approx$30\% of the
quiescent luminosity) could be due to a coronal flare of the
companion, since this would, again, be 3 orders of magnitude greater
than flares seen from other systems (see also R01).

\subsection{Spectral Analysis}
\label{sec:spectral}

Data were extracted, using {\em psextract}, into pulse-invariant (PI)
spectra. We binned all spectral data identically, as shown in their
best fit spectral figures (Figs ~\ref{fig:singlespec}a-d); the binning
was selected so that: (1) low-energy bins are wider than the detector
energy resolution; (2) the 0.45-0.60 keV bin integrates completely
across the O-edge in the detector, over which the detection efficiency
varies significantly; and (3) the high energy bins have comparable
signal-to-noise.  We analyzed the PI spectra using XSPEC v11.0.1
\cite{xspec}.

Our default spectral model is a H atmosphere spectrum \cite{zavlin96}
plus power-law and galactic absorption\footnote{Neutron star H
atmosphere emergent spectral models---and more complicated metallic
models---have also been calculated by \acite{zavlin96},
\acite{rajagopal96} and more recently by \acite{gaensicke01}.  The H
atmosphere models of these three references agree to within $\sim$few
per cent at energies relevant to the present work, 0.1-3 keV.}.  Other
spectral models (blackbody, Raymond-Smith plasma, pure power-law, disk
blackbody) for this source in quiescence have been examined elsewhere
(\rrr), using data from observation 1; note that in the present work,
the analyzed data extends down to 0.3 keV, while in the previous work,
data were not used below 0.5 keV.

We first fit the four spectra individually
(\S~\ref{sec:spectr-fits-indiv}), 
and followed this with combined fits of observations 1 and 2
(\S~\ref{sec:1and2}), and 3 and 4 (\S~\ref{sec:3and4}).  This was done
to address a systematic difference between the two groups of
observations.  We then attempted (\S~\ref{sec:joint-fitting}) joint fits
of all four observations to determine which spectral parameters
changed between them. 

\subsubsection{Spectral Fits to the Individual Observations}
\label{sec:spectr-fits-indiv}

We began by fitting all four spectra individually with an H atmosphere
model with galactic absorption and no power-law component.  While
obs. 1 and 2 are statistically acceptably fit (reduced chi-squared
statistic \chisqrnu=2.05/10 degrees of freedom, probability=0.03,
\chisqrnu=1.10/10 dof, prob=0.36 respectively), obs. 3 and 4 are not
(\chisqrnu=4.23/10 dof, prob=6\tee{-6}, \chisqrnu=6.8/10 dof,
prob=\ee{-10} respectively).  Visual examination of the spectra from obs. 3 and 4
show systematic excesses in counts above the model prediction at
energies $>$3 keV.  Such an excess has been observed previously from
this source \cite{campana98a} and other qNSs (see
\S~\ref{sec:powerlaw}), and is typically modeled as a power-law
component.  We therefore include this component in all four spectral
fits, to provide upper-limits on its flux in observations 1 and 2.
The resulting best-fit spectral parameters for the individual
observations are given in Table~\ref{tab:best}, and figures of the
best-fit models and data are in Figs.~\ref{fig:singlespec}a-d.

There is a systematic difference between the best-fit \nh\ value and
\rinfty\ values between the pairs of observations 1+2 and 3+4.  We
address this in the following sections, by comparing first the spectra
of observations 1 and 2, then the spectra of observations 3 and 4.

\subsubsection{Observations 1 and 2}
\label{sec:1and2}

As neither of these spectra statistically require a power-law
component, we investigate if the difference between them is due to
\nh, \kteffinfty, or \rinfty.  We jointly fit the two spectra with an
absorbed H atmosphere plus power-law, permitting 1 of the three
spectral parameters (\nh, \kteffinfty, \rinfty) to vary in turn
between the two observations.  We include a power-law component to
avoid systematic bias in the derived H atmosphere spectral parameters.
The resulting best-fit parameters and statistics are given in
Table~\ref{tab:specalla}.

The hypothesis of a variable \nh\ is rejected.  A variable \rinfty\ is
statistically acceptable, although not physically motivated in the
deep crustal heating model.  Finally, the hypothesis of a variable
\kteffinfty\ is statistically acceptable.  We therefore attribute the
difference between the spectra of these two observations to changes in
the \kteffinfty, which decreased from 125\ud{12}{9}~eV to
108\ud{11}{7}~eV over 81 days.  These uncertainties include covariance
with other spectral parameters (of which \nh\ and \rinfty are the
strongest).  When we hold all other parameters at their best-fit
values, the uncertainties are 125\ppm1~eV, and 108\ud{1}{2}~eV (1
$\sigma$).  Note that the best-fit \nhtt=0.43\ppm0.03 is consistent
with values found through other methods (Table~\ref{tab:nh}).

Attributing the variability to the magnitude of the power-law
component is statistically permitted by the data; however, in so
doing, the best-fit photon index is required to be steep (4.5\ppm0.2)
as to necessarily dominate the emission across the full \chandra\
energy band ($>$95\% of the 0.5-10 keV luminosity), the thermal
component is not required statistically (\rinfty$<$7.2 km at a
fiducial value of \kteffinfty=115 eV), and the value of \nh\ is a
factor of 2 above the optically implied value.  As we know of no
physical scenario in which such a steep power-law spectrum is
produced, and it would require the luminosity to be serendipitously
consistent with that of a NS with \rinfty$\approx$15 km at the
observed distance and \nh, we think this hypothesis unlikely.  We
therefore attribute the different intensities of observations 1 and 2
to a change in \kteffinfty.

\subsubsection{Observations 3 and 4: A Possible Counts Deficit in the
0.45-0.6 keV Bin}
\label{sec:3and4}

We jointly fit observations 3 and 4 with  the same absorbed H
atmosphere plus power-law spectrum.  A statistically acceptable fit is
found (Table~\ref{tab:specalla}).  Thus, \aql\ observations \#3 and \#4
are consistent with having the same spectrum and intensity.  The best
fit \nhtt\ (0.60\ppm0.05) is greater than that found by other methods
(Table~\ref{tab:nh}), including the value found from joint spectral
fitting from observations 1 and 2.  The value of \rinfty\ is
marginally larger than in observations 1 and 2 (46\ud{23}{11} km
vs. 17.2\ud{2.6}{3.6} km).  In addition, the value of \rinfty\ is also
larger than that found in a previous analysis of observation 1 alone
(\rrr), in which we had used counts with energies $>$0.5~keV, and that
found in analyses of \rosat\ and \asca\ data in which \nh\ was held
fixed at the optically implied value \cite{rutledge99}.

When we hold \nhtt\ and \rinfty\ fixed at the best-fit value from
observations 1 and 2, the best-fit spectrum is marginally acceptable
(\chisqrnu=1.71/25 dof, prob=0.015).  Examination of the spectral
model and data (Fig.~\ref{fig:34}) indicates a deficit in the
0.45--0.6 keV energy bin: the number of detected counts appears
significantly below that predicted by the best fit model.  When the
0.45--0.6~keV energy bin is removed from the fit, the new best-fit
model is significantly improved (\chisqrnu=0.76/23 dof; prob=0.78).
This suggests a deficit of counts in the 0.45--0.6~keV range in
observations 3 and 4 (largely, the latter).

The best-fit model (ignoring this energy bin, with \nhtt\ and \rinfty\
held fixed at the best-fit observation 1 and 2 values) predicts 20 and
25 counts in this bin for obs. 3 and 4 respectively (the difference in
counts is due to different integration times).  Only 14 and 10 counts
are observed in observations 3 and 4.  For an average countrate of 25
counts/bin, the Poisson probability of detecting $\leq$10 counts is
0.06\%.  Combining the two observations, the probability of detecting
$\leq$24 counts (with an average of 45 counts/bin) is 0.04\%.  No such
deficit is apparent in the spectra of observations 1 or 2.

\subsubsection{Joint Fitting}
\label{sec:joint-fitting}

To examine which spectral parameters are different between the four
observations, we jointly fit all four spectra simultaneously.  Because
of the similarity between observations 3 and 4, we treat their spectra
as being identical (i.e., all parameters for observations 3 and 4 are
treated as the same), so we are examining the difference among three
spectra, observations 1, 2 and 3+4.  Table~\ref{tab:specallb} contains a
summary of our findings.

We fit the data with the absorbed H atmosphere + power-law spectrum and
limited our fits to the 0.6--9.0~keV energy range, to avoid affects
associated with the deficit below 0.6 keV in observation 4.  As may be
surmised from the different countrates, using a single spectrum with all
five parameters (\nhtt, \rinfty, $\alpha$, \fxpl, and \kteffinfty) free
but the same for all 3 spectra, the best-fit is unacceptable
(\chisqrnu=5.52/43 dof; prob=\ee{-28}).  We then allowed each of the
spectral parameters to vary between observations.

\begin{enumerate}
	\item A fit with \kteffinfty\ allowed to vary is statistically
	acceptable, for values of \kteffinfty=121\ud{13}{6}~eV,
	105\ud{9}{6}~eV, and 110\ud{12}{5}~eV for observations 1, 2,
	and 3+4 respectively.  When we hold all other parameters fixed
	at their best-fit values (to exclude covariance in the
	uncertainties) the values are 121\ppm 1, 105\ppm 1, and
	110\ppm 1 eV.  

	\item When we permit only \nh\ to vary between the three
	spectra, the best-fit is unacceptable (\chisqrnu=3.05/41 dof;
	prob=2\tee{-10}).  Thus the changes between the spectra cannot
	be due only to a changing \nh.

	\item Although not physically motivated, when we permit only
	\rinfty\ to vary between the three spectra, we obtain an
	acceptable fit (\chisqrnu=1.13/41 dof; prob=0.26).  The range of
	acceptable \rinfty\ values is small: 12.5--17.6~km.

	\item We then fit the data with only \fxpl\ allowed to vary.
	The best fit was not statistically acceptable
	(\chisqrnu=1.60/41 dof; prob=0.0085).  When we permit both
	$\alpha$ and \fxpl\ to vary, the fit becomes acceptable; the
	changes between the power-law slopes are large, however, and
	the best-fit \rinfty\ is larger than expected from theory, and
	\nh\ is larger than that found through radio and optical
	observations.  It is unlikely, therefore, that a variable
	power-law component is responsible for the different spectra.

\end{enumerate}

The spectra of observations 1 and 2 are acceptably fit without a
power-law component, and the combination of the three spectra are
acceptably fit with only a changing \kteffinfty.  This motivated a fit
with the following parameters: no power-law component in observation 1
and 2; \nh\ held fixed at the radio measured value (\nhtt=0.34);
\rinfty\ held fixed at 13.0~km (a value which nearly all theoretical
EOSs can produce; \citenp{lattimer01}); and \kteffinfty\ varying.  The
best-fit is acceptable (\chisqrnu=1.30/43 dof; prob=0.11).  When we
now expand the energy range of this fit to include 0.3--0.6~keV,
however, the best fit becomes unacceptable (\chisqrnu=2.63/51 dof;
prob=2\tee{-9}).  Examination of the spectrum shows that the observed
countrates are systematically lower than the predicted countrates in
the 0.3--0.6~keV range.  When we let both \nh\ and \rinfty\ float
(with the same value in all three spectra), the best-fit spectral
model is still not statistically acceptable (\chisqrnu=1.62/49 dof;
prob=0.0038). This unacceptable spectral model is shown in
Fig.~\ref{fig:unaccept}.  The most discrepant point is the
0.45--0.6~keV bin from observation 4.  When this point is removed, a
new best-fit model is acceptable (\chisqrnu=1.26/48 dof; prob=0.105).

Since the poor fit to the full data set can be attributed to the
single low-energy bin in observation 4, one possibility is that \nh\
changed between observations 1+2 and observation 3+4.  If we permit
the values of \nh\ and \kteffinfty\ to change between observation 1, 2
and 3+4, however, the spectrum is still not acceptably fit
(\chisqrnu=1.66/47 dof; prob=3\tee{-3}); therefore, the deficit is not
explained by a changing \nh.

In summary, the differences in the three spectra are acceptably
explained as entirely due to a changing \kteffinfty\ of the surface
thermal emission.  The differences cannot be explained as entirely due
to either a changing power-law flux or to a variable \nh.  While a
combination of variability in $\alpha$ and \fxpl\ provides a
statistically acceptable fit, the required variation in the power-law
slope $\alpha$ is large and the resultant best-fit values of \rinfty\
and \nh\ are larger than we would \emph{a priori} expect. {\it We
therefore conclude that the simplest explanation for the variation in
the spectrum of \aql\ in quiescence is due to a change in \kteffinfty\
of a pure H atmosphere.
}

\subsection{Re-examination of the Counts Deficit in Observation 4}

We now turn our attention to the apparent deficit in counts in the
0.45--0.6 keV bin of observation 4.  In particular, we examine whether
it is possible to explain the deficit by some instrumental or analysis
effect that is unrelated to the NS emission.  

When we exchange the response matrix from obs. 4 to that from obs. 1,
there is no effect on the spectral fit (the focal plane temperatures
were the same in all observations, and the X-ray source was focussed
at the same location on the S3 chip).  The deficit is therefore not
caused by a change in the calibrated spectral response particular to
obs. 4.

What if the effective area of the 0.45-0.6 keV bin is simply different
from that represented in CALDB v2.9?  Could we simply change the
expected number of counts (on the basis of the best-fit spectrum and
the calibrated response) by the same constant factor, which would make
all observed 0.45-0.6 keV countrate consistent with the expected
number?  The calibrated response is based on detailed knowledge of the
telescope+detector system.  We took the response matrix to be
completely diagonal, such that a photon of energy $E_n$ would be
detected as a count with energy $E_n$.  While the system response is
considerably more complicated, with significant non-diagonal elements,
this zeroth order approximation can reveal if change in the 0.45-0.6
keV effective area can account for the observed discrepancy, or if
higher order changes would be necessary.

In Table~\ref{tab:counts}, we list the number of counts detected in
the 0.45-0.6 keV energy bin $x_i$ during each observation $i$, and the
number of counts predicted to be in the bin ($\mu_i$) by our best-fit
spectral model (in which the 0.45--0.6 keV bin was ignored in
observation 4).  We then calculated a probability for the factor ($f$)
by which we change the diagonal response,
\[
Prob(f) = \Pi_{i=1}^4  {P_{\rm Poisson}(f\times\mu_i, x_i)}
\]
where $P_{\rm Poisson}(\mu, x)$ is the Poisson probability of
observing $x_i$ counts when $\mu_i$ is the expected average
realization.  We find, for our observed and predicted values of $x_i$
and $\mu_i$, that $Prob(f)$ is maximal when $f=0.75$ (that is, the
observed counts would be most consistent with an effective area in the
0.45-0.6 keV bin which is 75\% of the telescope+detector effective
area represented in CALDBv2.9).  We performed a Monte Carlo
simulation, with 4 Poisson deviated counts, with mean counts of
$f\times\mu_i$, calculating $Prob_{\rm Monte-Carlo}(f=0.75)$ for
\ee{6} such realizations.  We found that we would produce a value of
$Prob_{\rm Monte Carlo}(f=0.75)\leq Prob(f=0.75)$ 2.0\% of the time
(that is, even the most likely value of $f$ would produce the
distribution of $x_i$ observed only 2\% of the time).  We thus
marginally exclude the counts deficit as due to a different effective
area in the 0.45-0.6 keV bin from that in CALDB 2.9.  To attribute the
observed discrepancy to a response inaccuracy, there must either be a
significant difference between a non-diagonal element as modelled and
its true value, or the diagonal element response in the 0.45--0.6 keV
bin (and only that bin) may be time dependent.

We also examined the spectrum in PHA space as well as PI space, and
find the same deficit of counts in observation 4.  A statistically
anomalous background subtraction (i.e, more than an average number of
background counts in this spectrum) is not responsible for the deficit
either; when we neglect the background entirely (which we estimate to
be $\sim$9 counts of the 1137 counts observed in this spectrum), the
deficit remains.

We therefore marginally exclude the deficit of counts to response
matrix uncertainty under a diagonal assumption, and exclude either
anomalous background, or binning in PI space vs. PHA space.  The
deficit could be due to non-diagonal elements in the redistribution
matrix being different than those in CALDB~v2.9, or to an unknown
time-dependency in the diagonal elements of the redistribution
matrix. 

\subsection{Time Dependent Response Below 0.7 keV in ACIS-S3}
\label{sec:resp}

When this work was largely complete, it was reported that there is a
time-dependency in the response of the ACIS chips below 0.7 keV
\footnote{http://asc.harvard.edu/cal/Links/Acis/acis/Cal\_projects/index.html},
which appears as a decrease in effective area, the magnitude of which
has increased with time since launch.  The magnitude, exact time
dependency, and energy dependency are not fully known at present, and
it is beyond the scope of the present work to provide a full time- and
energy- dependent re-calibration effort.  However, the discussed
possible causes (deposition of contaminants on the CCDs or the optical
blocking filter) suggest the effect would be monotonicaly increasing
with time, perhaps at a constant rate. 
 
We examined briefly the magnitude of this effect, to determine its
possible effects on our conclusions.  We used two observations of the
spectrally soft super-nova remnant SNR E0102-72.3, taken 2000 May 28
(ObsID 141, $\sim$6 months prior to Obs. 1 of Aql X-1) and 2001 Dec 6
(ObsID 2844; $\sim$8 months after Obs. 4 of Aql X-1). These were taken
with the same focal plane temperature ($-120F$) but at different
locations on the ACIS-S3 chip (141 was 2.2\arcmin\ off-axis, while
2844 was only 1\arcmin\ off axis).  We extracted data within
39\arcsec\ of the SNR center, and produced pulse-invarint (PI) spectra
with resolution (10-15 eV) greater than the intrinsic energy
resolution of ACIS-S3 between 0.3-1.1 keV (120-140 eV).  We 
modelled the first observation with a model meant to be parametric
({\tt wabs * (c6vmekl + gauss)}), which with a best fit of
\chisqrnu=3.30 for 32 dof was not statistically acceptable;  however
the most deviant PI bins were not systematically offset from the model
(that is, the scatter appeared random as a function of energy), and
were at most 3$\sigma$ from the model. This is adequate for our
purpose of investigating systematic response differences at the energy
resolution of the detector. 

We compared the second spectrum obtained 557 days later with the
best-fit model of the first spectrum (see Fig.~\ref{fig:e0102}).  As
can be seen from the figure, the response appears to be diminished by
$\sim$50\% between 0.3-0.5 keV, with the magnitude of the discrepancy
decreasing between 0.5-0.7 keV.  Thus, from our comparison, we see an
apparent change in the detector response, which may be modelled as an
energy-dependent decrease in the effective area below 0.7 keV.

We investigate the effect of this on our conclusions regarding the
broad-band spectrum, by using data only in the energy range 1-10 keV.
First, as can be seen from Table~\ref{tab:obs}, the 1.0-2.5 and 2.5-8
keV countrates vary between observations, and so cannot be explained
as due to the $<$1 keV response.  Joint spectral fits using all four
observations still find that the variation cannot be explained as due
to a change in \nh\ or power-law component normalization alone (both
with values of \chisqrnu\ with corresponding probabilities of
$<$\ee{-3}).  Since we are simply removing data from consideration,
all those combinations of parameters which produced acceptable
spectral fits using 0.3-10 keV still produce acceptable fits using
1-10 keV, but with larger error bars.  Thus, the conclusion that the
observation-to-observation variability is explained by a change in
\kteffinfty\ remains.  We expect the time-dependent calibration of the
sub-1 keV range, when it is fully known, will change the resulting
spectral parameters we have measured here, but not our qualitative
conclusions.  

Finally, as regards the apparent counts deficit, the change in
effective area does affect the energy range of interest (0.45-0.6
keV).  We performed the following check. First, using the E0102-72.3
observations, we included a multiplicative {\tt spline} model in XSPEC
to account for the change in calibration; this was done by first
finding the best-fit parametric model to the May '00 observation only,
and using the same parameters held frozen for the Dec '01 data, but
including a {\tt spline} model, with Estart=0.3 keV, Eend=0.7 keV,
Yend=1.0, YPend=0.0, and the values of Ystart and YPstart permitted to
vary.  The best-fit spectrum was unacceptable (\chisqrnu=15.3, 52
dof), although the gross change in response was well accounted for,
with best-fit values of Ystart=1.5 and YPstart=$-20$.  We then refit
Obs. 1-4 with the spectral model which had been previously
unacceptable due to the counts deficit (\kteffinfty\ varies between
observations, \nh\ and \rinfty\ permitted to float, and a power-law
spectral component for Obs. 3 and 4 only), but including the spline
component for Obs 3 and 4.  This models the change in response as
essentially instantaneous between observations 2 and 3.  The best-fit
joint spectrum is statistically acceptable, and the deficit is no
longer significant (\chisqrnu=1.23 / 49 dof; prob=0.13).  The accuracy
of this correction, however, depends in detail on the time-dependence
of the change in response, and whether or not it can be modelled as a
change in effective area as a function of energy.  This requires much
more detailed modelling of the response than we can do here.  We
conclude, however, that the deficit of counts can be due entirely to
the change in detector response; this requires confirmation through
detailed study and modelling on the time- and energy-dependence of the
sub-0.7 keV detector response.

\section{Discussion and Conclusions}
\label{sec:con}

Following an outburst of \aql\ during November 2000, we took four 
{\em Chandra}/ACIS-S snapshots over a span of 5 months.  With this 
series of spectra, we find the following.

\begin{enumerate}
\item All four spectra are acceptably fit with H photosphere models, 
with radii consistent with that of a NS, at the distance and reddening
of Aql X-1.  In two of the four spectra, an additional power-law
component is required to acceptably fit the spectrum at high energies
($>$3 keV), as has been found previously for this and other qNSs.

    \item The intensity decreased by 50\% over the first three months
    following the outburst.  It then increased by 35\% over the next
    month and then remained constant for the last month spanned by our
    observations.  The change in the spectrum between the observations
    cannot be explained exclusively as a change in the power-law
    component, nor in the value of \nh.  A change in the photospheric
    effective temperature is required to account for the change.

    \item No short timescale ($<10^{4}$~s) variability was found in
    the first three observations; the fourth had significant
    variability (32\ud{8}{6}\% rms, 0.2-10 keV) on timescales
    \ee{3}--\ee{4} sec.  As the power-law component contributes a
    smaller fraction ($\sim$12\%) to the total counts, this magnitude
    in variability cannot be attributed to the power-law component.

    \item A deficit of counts in the $0.45$--$0.6$~keV band was noted
    during observation 4.  This deficit was not present in the other
    three observations.  Follow-up spectroscopy and detailed study of
    the change in the detector response is required to determine if
    the deficit is intrinsic to the source.

    \item A best-fit spectral model, excluding the 0.45-0.6 keV bin in
    observation 4, gives a value of \nhtt=0.42\ud{0.02}{0.03} --
    consistent with the optical and radio implied value; the best fit
    \rinfty=15.9\ud{0.8}{2.9} km (neglecting the 20\% uncertainty in
    the 5 kpc distance), consistent with the radius of a NS for most
    proposed NS EOS \cite{lattimer01}.
\end{enumerate}

The increase in temperature between observations 2 to 3, if significant,
is interesting.  While accretion can clearly affect the thermal spectrum
(e.g., Zampieri et al.~1995), it is also possible that differential
sedimentation of ions in the atmosphere can change the surface effective
temperature by modifying the opacity at intermediate densities,
$10^{5}\mathrm{\;g\;cm^{-3}} \lesssim\rho\lesssim
10^{8}\mathrm{\;g\;cm^{-3}}$, \cite{brown02}.  Further
detailed calculations are required to determine the amplitude and
timescale of this effect.

At present, there is no known mechanism associated with crustal
heating that can produce the short-timescale ($\approxlt$\ee{4} sec)
variability we observe here. The minimum timescale for post-outburst
intensity variability in quiescence from deep crustal heating is days
to months (the thermal diffusion timescale from the depth of heat
deposit), and should scale roughly as $\delta t$, where $\delta t$ is
the time since a (short) outburst ends \cite{ushomirsky01}.  For NSs
with ``normal'' cores, the magnitude of variability is a few per cent;
if more exotic neutrino cooling takes place in the core, however, the
magnitude of variability in quiescence can be a factor of ten or more.
In the absence of variability due to post-outburst crustal cooling,
the only other timescale for variability is the timescale of core
cooling, which is typically \ee{5}--\ee{6} yr.
In contrast, intensity variability due to accretion onto the compact
object can in principle take place on the dynamical timescale, seconds
or shorter.  However, while the observation is in qualitative
agreement with accretion onto the compact object, it remains a major
puzzle why the magnitude of luminosity should be quantitatively
comparable to the luminosity predicted previously from deep crustal
heating \rrr.  The absence of a quantitative prediction for the
magnitude of accretion onto the compact object in quiescence further
complicates comparison between this observation and an accretion
theory.  For the present, we conclude that the intensity variability
on $<$\ee{4} sec timescale can be explained by accretion onto the
compact object, but the implied coincidence of the accretion
luminosity being comparable to the predicted deep crustal heating
luminosity, as found by \rrr, remains a challenge to theory.

\subsection{Uncertainties in the Measurement of $R$, and Prospects for a Useful Constraint on the EOS}

Throughout this paper, we use the word "radius" to mean the measured
value \rinfty, the effective emission area size observed from an
infinite distance.  In practice, what we measure using spectral
analysis is the angular size of the source \rinfty/$D$, where $D$ is
the distance to the NS.  We assume $D=$5kpc, a value which is at
present uncertain by \ppm20\% (\rrr); we neglect this uncertainty
throughout the present work, addressing only the statistical and
spectral uncertainty, which can be greatly improved upon with higher
signal-to-noise X-ray spectra.  The distance to \aql\ can be
accurately measured in the future using the Space Interferometry
Mission (SIM; \citenp{sim}), which can achieve \ppm2\% accuracy at 5
kpc for an object with $V<17m$ (Aql X-1 obtains $V=14.8$ in outburst ;
\citenp{liu01}).  Such precise measurement will effectively eliminate
distance as an uncertainty in \rinfty.  Indeed, using spectral
analyses of qNSs in globular clusters
\cite{grindlay01,rutledge01c,intzand01}, the distances to which can
be measured with \ppm2\% precision using Hipparcos dwarfs
\cite{carretta00}, is one approach being pursused to obtain accurate
\rinfty\ measurements prior to the launch of SIM.

We have assumed throughout this work that \rinfty\ does not change.
In the deep crustal heating scenario (BBR98), luminosity originates
from the NS core, and the NS star has a \kteffinfty\ which is
isotropic at the photosphere, which we assume throughout this paper.
This scenario, however, is not entirely consistent with the short-term
(seconds-months) variability which is observed.  If the photosphere
were to have temperature anisotropies -- for example, which may occur
if the NS were powered exclusively by non-spherically symmetric
accretion -- this would change the physical meaning of \rinfty, which
would then dependent on the details of the photospheric temperature
distribution, as described for accreting NSs at much higher
luminosities than are relevant here \cite{psaltis00}.

The value of \rinfty\ is related to the physical radius $R$ by
\rinfty=$R/\sqrt{1 - 2GM/(Rc^2)}$, where $G$ is the graviational
constant, $M$ is the NS mass, and $c$ is the speed of light.  It is
the value $R$ which has recently been shown to place a useful
constraint on the NS EOS \cite{lattimer01}, if $R$ is measured to
\approxlt 10\% accuracy.  The value $R$ cannot be derived from
\rinfty\ in the absence of a known NS mass $M$.  If one assumes a
range of values between 0.8-2.8\msun, this introduces an additional
systematic uncertainty of $\sim$50\% in $R$, which is greater than
that needed to usefully constrain the EOS in the prescription of
\acite{lattimer01}.  Thus, precise measurements of \rinfty\ will not
usefully constrain the NS EOS in the absence of a measured NS mass or
photospheric redshift; however, if a phenomenological relationship
between \rinfty\ and the pressure at nuclear density for a range of
EOSs can be found, as found for $R$ by \acite{lattimer01}, it may be
that \rinfty\ will usefully constrain the NS EOS.  This possibility is
presently under examination.

\subsection{The Power Law Component}
\label{sec:powerlaw}

A power-law spectral component is required to fit the spectrum above 3
keV in 2 of the four observations, and it is consistent with being the
same magnitude and slope in all four observations.

A high energy power-law tail has now been observed to dominate the
X-ray spectrum above 2-3 keV in five qNSs: Aql X-1, \cenx4
\cite{asai96b,campana00,rutledge00}, KS 1731-260
\cite{wijnands01b,rutledge01d}, NGC 6440 \cite{intzand01} and
4U~2129+47 \cite{nowak02}.  In all cases, this component comprises
$\sim$10-40\% of the 0.5-10 keV luminosity.  Importantly, in all those
power-law components which have been detected, the slope appears to be
either inverted or flat.  A cutoff at higher photon energies (above
3~keV) is required for the luminosity not to diverge.

It has previously been suggested that the component may be due to
accretion onto the NS magnetosphere \cite{campana98b}; however, no
detailed {\em a priori} model (including an energy budget, estimation
of the accretion rate onto the compact object or magnetosphere)
exists.  If the power-law emission were powered by accretion onto the
magnetosphere, however, it would be coincidental that five different
sources (each having different $B$ fields, spin rates, and perhaps
accretion rates onto the magnetosphere) would have comparable
luminosity ratios of the power-law component to the thermal component,
which originates from the NS surface.  The same coincidence would
apply to non-thermal emission from pulsar shocks as well
\cite{tavani91,stella94,menou99}. 

One possible suggestion is that both the thermal and power-law
spectral components are powered by low-level accretion. Using the
assumption of spherical flow onto an unmagnetized, nonrotating neutron
star, and assuming that only Coulomb forces acted to decelerate the
infalling protons, \acite{zampieri95} showed that no similarly hard
spectral component was expected at accretion rates comparable to that
required to power the quiescent thermal emission.  \acite{deufel01}
similarly found that, for accretion via an advective flow with
virialized protons, the emergent spectrum is slightly harder than
thermal, though the hardened component still decreases in $\nu F_\nu$.
It is therefore unclear that the emergent spectrum of a low $\dot{M}$
NS would produce a hard tail such as observed here.

\subsection{A Deficit of Counts in the 0.45-0.6 keV Bin in Observations 3 and 4}

We found that, compared with a spectral model interpolation in the
0.45-0.6 keV energy range, there is a deficit of counts significant at
the 99.96\% level, during observations 3 and 4.  Such a deficit is
not present in observations 1 or 2.  There are at least a few
possible explanations for this. 

It may simply be that the deficit is a statistical anomaly, which will
not be confirmed with deeper spectroscopy.  We may be able to address
this possibility with our approved 30ksec AO-1 GO observation of \aql\
with XMM-Newton.

A simple decrease in the effective area in the 0.45-0.6 keV energy
bin, by a factor of 0.75, is marginally excluded (2\% probability) as
the source of the counts deficit. The deficit may be due, however, to
non-diagonal elements of the photon energy redistribution matrix which
are different from those in the present calibration model. Moreover, a
recently reported time-dependence in the ACIS response below 0.7 keV
can expalain the entire deficit, when the time dependence of the
change in response is modelled as having taken place between Feb and
April 2001 (observations 2 and 4).  Until this response is properly
modelled, it cannot be neither fully implicated nor exonerated as
being responsible for the counts deficit.  Thus, an unaccounted-for
discrepancy in the detector response remains a possible explanation.

Finally, it may be that the deficit of counts is due to the presence
of metals.  This would be consistent with the observational scenario
expected to produce such features \cite{brown98}, in which active
accretion onto the NS surface replenishes the surface metallicity. The
detected intensity variability (32\ud{8}{6}\% rms) argues for active
accretion during this observation as well (such variability may also
be present during observation 3); and the luminosity of \aql\ during
these observations is consistent with this scenario as well.

  We show in Fig.~\ref{fig:slava} the emergent spectra of a
solar-metallicity atmosphere from a NS with the effective temperature
of \aql, rotating at 550 Hz \cite{zhang98b}, at three different
viewing angles with respect to the rotation axis (the orbital plane of
\aql, is $\alpha>36^\circ$).  The absorption features due to metallic
absorption lines are apparent in the pole-on ($\alpha=0^\circ$) spectral
model.  The most prominent feature, smeared by NS rotation, is near
0.9 keV, which is a mixture of OVII absorption, plus other metal
absorption lines.  In a rough approximation: to redshift this feature
to 0.45-0.6 keV requires a redshift $1+z =
g_r^{-1}=1/\sqrt{1-2GM/(Rc^2)}$=1.5-2.0, which, for a NS of \rinfty=$R
g_r^{-1}$=15.9 km, implies a coordinate radius $R=8-10.5$ km, and a
mass $M\approx$2.0\msun.  This is a mass/radius estimate of scale, which
shows consistency with theoretical expectations, but is not a precise
estimate due to the large uncertainty in the redshift factor, in
addition to the systematic uncertainty in the correct identification
and reality of the feature in question.

If future observations confirm the presence of this absorption line in
\aql\ in quiescence, it can be exploited. By measuring the redshift of
such lines ($g_r^{-1}$) simultaneously with the emission area from the
spectrum ($\propto R^2\, g_r^{-2})$ the
NS mass and radius can be determined, providing a strong constraint on
the equation of state of dense matter \cite{lattimer01}.

\acknowledgements

We acknowledge and thank an anonymous referee, whose useful comments
 improved the manuscript.  We thank Ira Wasserman for discussions and
 Divas Sanwal for critically important discussions regarding the
 time-dependent soft energy response of ACIS.  The authors are
 grateful to the \chandra\ Observatory team for producing this
 exquisite observatory.  This research was partially supported by the
 National Science Foundation under Grant No. PHY99-07949 and by NASA
 through grant NAG 5-8658, NAG 5-10855 and the \chandra\ Guest
 Observer program through grant NAS GO0-1112B.  L. B. is a Cottrell
 Scholar of the Research Corporation.  E. F. B. acknowledges support
 from an Enrico Fermi Fellowship.

\clearpage

\pagestyle{empty}
\begin{figure}[htb]
\caption{ \label{fig:lc} Lightcurve of Aql X-1, outburst through
quiescence.  \rxte/ASM points (squares) used the ASM countrates,
assuming 1 ASM c/s=4.4\tee{-10} \cgsflux\ (0.5-10 keV; appropriate for
a $kT=5$ keV thermal bremsstrahlung spectrum corrected for
absorption), and the ACIS-S points (triangles) assume the spectrum for
observations 3 and 4 (1 ACIS-S c/s=1.0\tee{-11} \cgsflux, 0.5-10 keV).
Only ASM detections of $>4\sigma$ are used. The uncertainty in
luminosity due to the countrate is shown; the uncertainty in
luminosity does not include the systematic uncertainty in the
bolometric correction.}
\end{figure}
\nocite{leahy83}

\pagestyle{empty}
\begin{figure}[htb]
\caption{ \label{fig:pds} PDS of observations 1-4. Points are data,
normalized according to Leahy \etal (1983).  Error bars and
upper-limits are 1$\sigma$.  Solid lines are the best-fit broad-band
models (see text). RMS variability model values are 0.0001-1 Hz, for
all counts data detected (roughly, 0.2-10 keV).  Observations 1-3 give
only upper-limits for variability, while in observation there is a
clear detection of variability, with 32\ud{8}{6}\% rms.  }
\end{figure}
\nocite{leahy83}

\pagestyle{empty}
\begin{figure}[htb]
\caption{ \label{fig:obs4lc} Lightcurve of observation 4.
}
\end{figure}

\pagestyle{empty}
\begin{figure}[htb]
\caption{ \label{fig:singlespec} Best fit spectra for observations
1--4 (parameters are in Table~\ref{tab:best}). Crosses are the observed data.  The dashed line is the thermal component, 
the dashed-dotted line is the power-law component, and the solid line is the summed spectral model; the spectral models 
have been corrected for galactic absorption.   {\bf (a)} Obs. 1.  {\bf (b)} Obs. 2. 
{\bf (c)} Obs. 3.  {\bf (d)} Obs. 4. 
}
\end{figure}

\pagestyle{empty}
\begin{figure}[htb]
\caption{ \label{fig:34} (Top Panel) The observed spectra for observations 3 (open
squares) and 4 (open circles), and the best-fit model in which the
column density is held fixed at \nhtt=0.38 -- the best-fit value from
observations 1 and 2, as well as comparable to the optically implied
value (\nhtt=0.35).  The spectrum is statistically unacceptable, due
to the deficit of counts in the 0.45-0.6 keV bin.  (Bottom Panel) The
residuals $\chi$=(data-model)/uncertainty. }
\end{figure}

\pagestyle{empty}
\begin{figure}[htb]
\caption{ \label{fig:unaccept} (Top Panel) Observations 1--4, with an
(unacceptable) best spectral fit (see text), due to the low countrate
in the 0.45-0.6 keV bin of observation 4. Observation 1 (stars), 2
(triangles), 3 (squares), and 4 (circles).  (Bottom Panel) Residuals
$\chi$=(data-model)/uncertainty.
}
\end{figure}

\pagestyle{empty}
\begin{figure}[htb]
\caption{ \label{fig:slava} The emergent (top of the atmosphere --
that is, un-redshifted) X-ray spectrum for a NS atmosphere (1.4\msun,
10km) of solar metallicity including the effects of a NS rotating at
550 Hz, as observed from \aql, at three viewing angles ($\alpha$)
relative to the direction of NS rotation.  The solid line, in which
absorption features are apparent, is for an angle $\alpha=0$ of the NS
rotation axis relative to the line of sight.  The broken line is for
$\alpha=30^\circ$, and the smooth solid curve us for
$\alpha=90^\circ$.  $Z$ indicates the percentage of metals, assuming
the mixture of Grevesse and Noels (1993).  For \aql\ the best estimate
of the rotational inclination is $\alpha>36^\circ$.}
\end{figure}

\nocite{grevesse93}

\pagestyle{empty}
\begin{figure}[htb]
\caption{ \label{fig:e0102} {\bf Top Panel}: Comparison of two X-ray
spectra for the soft SNR E0102-72.3 between 0.3 and 1.1 keV, taken 5
May 2000 (crosses) and 6 Dec 2001 (boxes).  A parametric model (solid
line) is fit to the the May '00 observation.  The Dec '01 spectrum is
systematically below the May '00 at energies 0.3-0.7 keV.  ({\bf
Bottom panel}) Ratio of the best-fit May '00 parametric spectral model
to the May '00 (crosses) and Dec '01 (boxes) data.  The systematic
difference in spectra shows a strong energy dependence: approximately
50\% near 0.3-0.5 keV, increasing to consistency (100\%) near 0.7 kev.
Comparison of the increases and decreases in the PI spectra also
indicate a possible gain shift between the two observations, by
$\sim$10 eV.  }
\end{figure}

\clearpage
\pagestyle{empty}
\begin{figure}[htb]
\PSbox{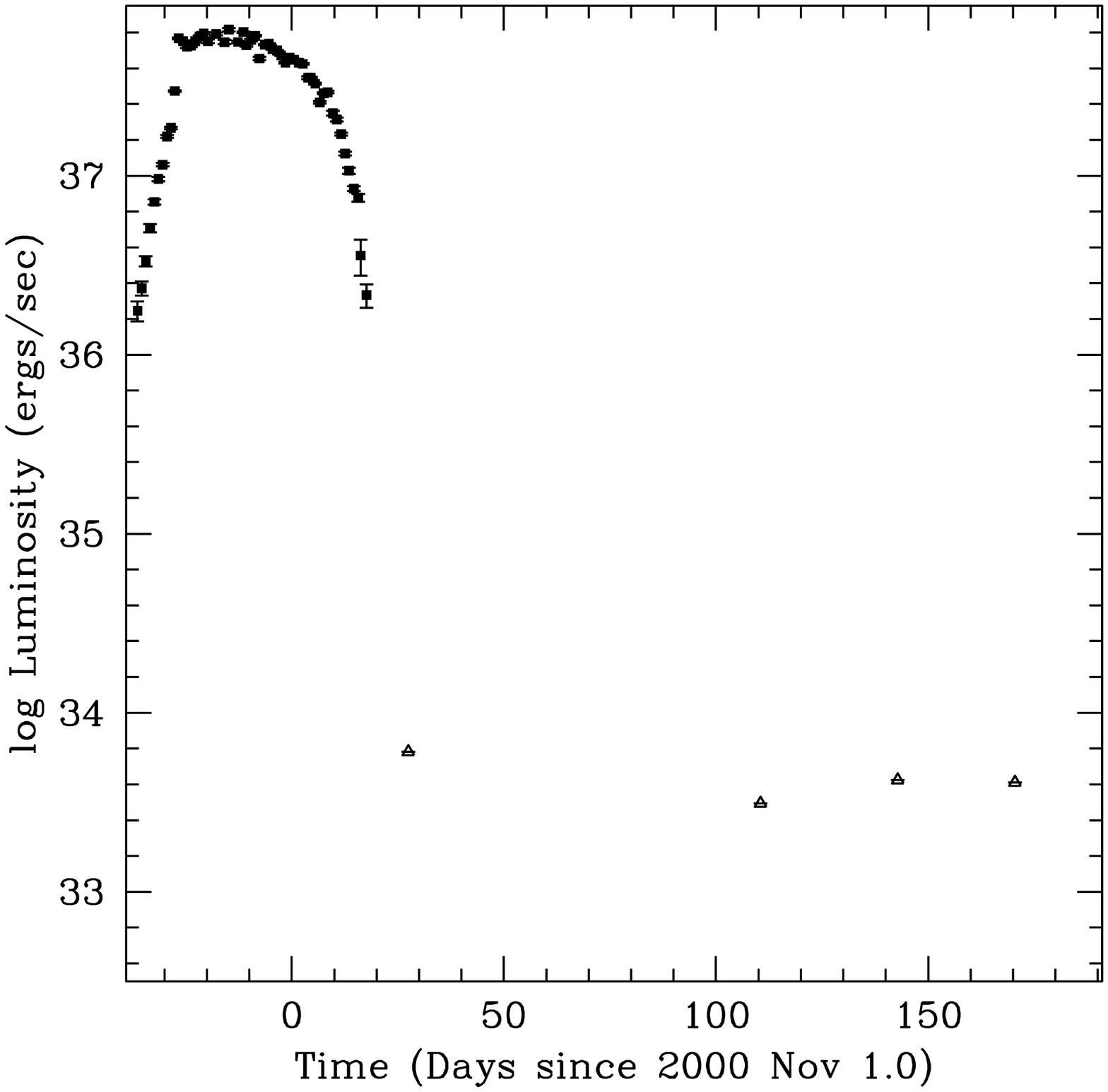 hoffset=-80 voffset=-80}{14.7cm}{21.5cm}
\FigNum{\ref{fig:lc}}
\end{figure}

\clearpage
\pagestyle{empty}
\begin{figure}[htb]
\PSbox{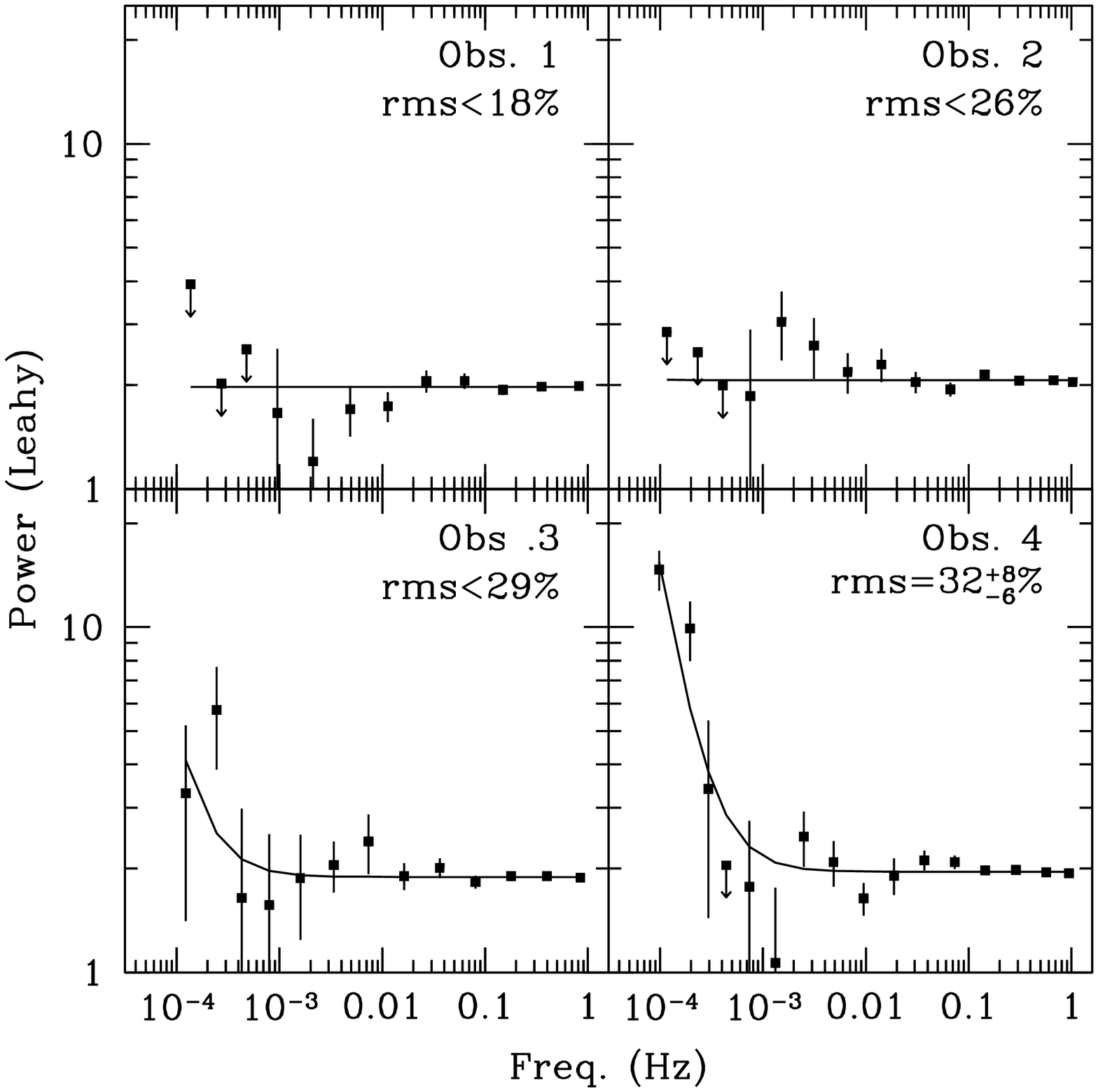 hoffset=-80 voffset=-80}{14.7cm}{21.5cm}
\FigNum{\ref{fig:pds}}
\end{figure}

\clearpage
\pagestyle{empty}
\begin{figure}[htb]
\PSbox{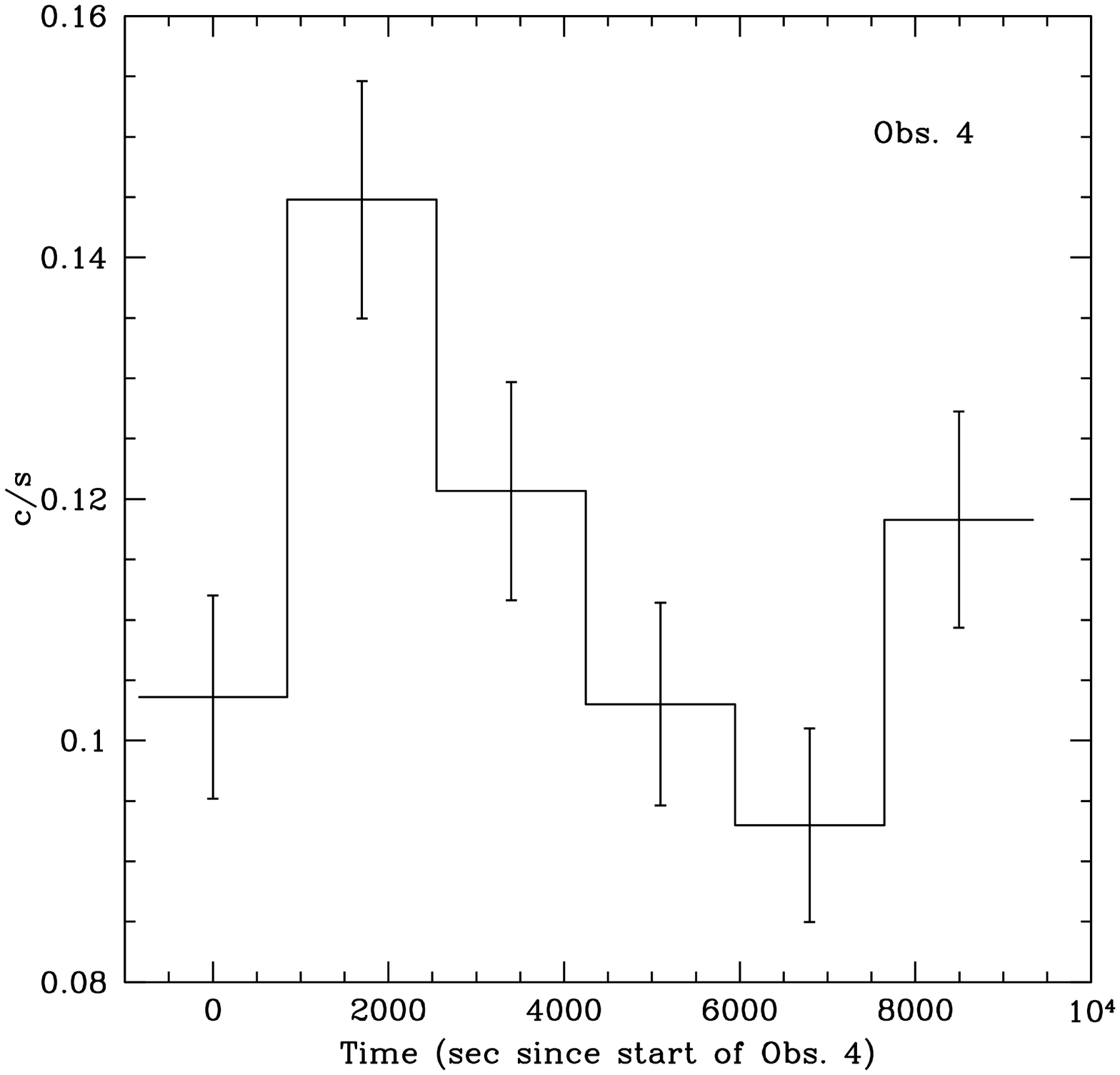 hoffset=-80 voffset=-80}{14.7cm}{21.5cm}
\FigNum{\ref{fig:obs4lc}}
\end{figure}

\clearpage
\pagestyle{empty}
\begin{figure}[htb]
\PSbox{f4a.ps hoffset=-80 voffset=-80}{14.7cm}{21.5cm}
\FigNum{\ref{fig:singlespec}a}
\end{figure}

\clearpage
\pagestyle{empty}
\begin{figure}[htb]
\PSbox{f4b.ps hoffset=-80 voffset=-80}{14.7cm}{21.5cm}
\FigNum{\ref{fig:singlespec}b}
\end{figure}

\clearpage
\pagestyle{empty}
\begin{figure}[htb]
\PSbox{f4c.ps hoffset=-80 voffset=-80}{14.7cm}{21.5cm}
\FigNum{\ref{fig:singlespec}c}
\end{figure}

\clearpage
\pagestyle{empty}
\begin{figure}[htb]
\PSbox{f4d.ps hoffset=-80 voffset=-80}{14.7cm}{21.5cm}
\FigNum{\ref{fig:singlespec}d}
\end{figure}

\clearpage
\pagestyle{empty}
\begin{figure}[htb]
\PSbox{f5.ps hoffset=-80 voffset=-80}{14.7cm}{21.5cm}
\FigNum{\ref{fig:34}}
\end{figure}


\clearpage
\pagestyle{empty}
\begin{figure}[htb]
\PSbox{f7.ps  hoffset=-80 voffset=-80}{14.7cm}{21.5cm}
\FigNum{\ref{fig:unaccept}}
\end{figure}

\clearpage
\pagestyle{empty}
\begin{figure}[htb]
\PSbox{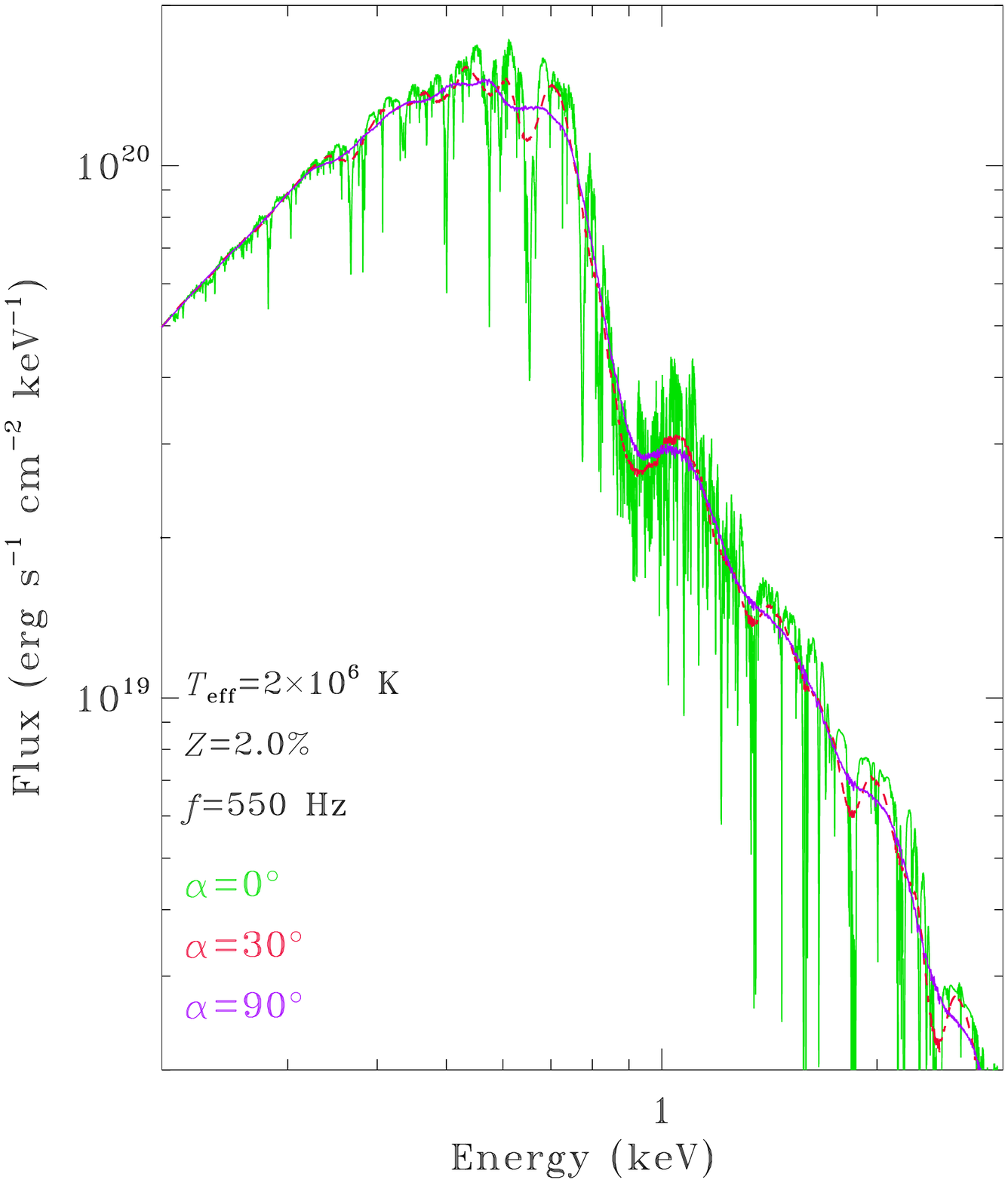  hoffset=-80 voffset=-80}{14.7cm}{21.5cm}
\FigNum{\ref{fig:slava}}
\end{figure}
\clearpage

\clearpage
\pagestyle{empty}
\begin{figure}[htb]
\PSbox{f9.ps  hoffset=-80 voffset=-80}{14.7cm}{21.5cm}
\FigNum{\ref{fig:e0102}}
\end{figure}
\clearpage

\begin{deluxetable}{cccccccc}
\scriptsize
\tablecaption{\label{tab:obs} Log of \chandra/ACIS-S Observations}
\tablewidth{18cm}
\tablehead{
\colhead{} & 
\colhead{Start Time} &
\colhead{Exposure} &
\colhead{c/ksec } &
\colhead{c/ksec } &
\colhead{c/ksec } &
\colhead{c/ksec } &
\colhead{Orbital Phase$^a$}\\
\colhead{Obs \#} & 
\colhead{(TT)} &
\colhead{(sec)} &
\colhead{(0.5-10 keV)} &
\colhead{(0.3-1.0 keV)} &
\colhead{(1.0-2.5 keV)} &
\colhead{(2.5-8 keV)}  &
\colhead{$\phi_{\rm orb}$}
}
\startdata
1	& 2000-11-28 10:52:39		&6628 &	183\ppm5& 56\ppm3 	& 120\ppm4      &8.3\ppm1.1&0.02--0.15 (\ppm0.01)	\\
2	& 2001-02-19 11:25:52		&7787 &	94\ppm3	& 32\ppm2	& 60\ppm3       &3.9\ppm0.7&0.20--0.36 (\ppm0.02) 	\\
3	& 2001-03-23 20:09:11		&7390 &	127\ppm4& 39\ppm3	& 81\ppm3       &8.2\ppm1.1&0.19--0.34 (\ppm0.02)	\\
4	& 2001-04-20 11:53:32		&9245 &	123\ppm4& 38\ppm2	& 76\ppm3       &9.9\ppm1.0&0.22--0.39 (\ppm0.02)	\\
\enddata 
\tablecomments{$^a$ Orbital phase relative to minimum light (inferior conjunction of the secondary), ephemeris from \acite{garcia99}, orbital period from \acite{chev98}}
\end{deluxetable}


\begin{deluxetable}{cccc}
\tablecaption{\label{tab:var} Intensity Variability }
\tablewidth{8cm}
\tablehead{
\colhead{Obs. \#} &
\colhead{$\alpha$} &
\colhead{Total Counts} &
\colhead{RMS (\%)} 
}
\startdata
1	&	(1.8)	&	1237	& $<$18	\\
2	&	(1.8)	&	755	& $<$26	\\
3	&	(1.8)	&	956	& $<$29	\\
4	&	1.8\ppm0.4&     1161    & 32\ud{8}{6}	\\
\enddata 
\tablecomments{Values in parenthesis are fixed. RMS is percentage root mean square variability, integrated 
between 0.0001-1 Hz, in excess of the Poisson level; error bars are 1$\sigma$, upper-limits are 3$\sigma$.}
\end{deluxetable}

\begin{deluxetable}{ccccccccc}
\scriptsize
\tablecaption{\label{tab:best} \chandra\ Best-Fit Spectral Parameters for
Individual Observations}
\tablehead{
\colhead{} & 
\colhead{} & 
\colhead{\kteffinfty} &  
\colhead{\rinfty}  & 
\colhead{} & 
\colhead{$N_{pl}$} & 
\colhead{$f_{pl}$} & 
\colhead{Flux$^a$} & 
\colhead{} \\
\colhead{Obs. \#} & 
\colhead{\nhtt} & 
\colhead{(eV)} &  
\colhead{(km/(D/ 5 kpc))}  & 
\colhead{$\alpha$} & 
\colhead{()} & 
\colhead{(\%)} &
\colhead{(0.5-10 keV)} &
\colhead{\chisqrnu\ (prob)} 
}
\startdata
1	&  0.46\ppm0.04			& 118\ud{8}{6} 	& 20\ud{5}{3}		& (1.0)		   &  $<$2\tee{-5}&  $<$17\%	& 18	&  0.83/10 (0.07)	\\
2	&  0.37\ppm0.06			& 120\ppm20 	& 13\ud{6}{4}		& (1.0)		   &  $<$1.2\tee{-5}&  $<$22\%	& 8	&  0.24/9  (0.99)	\\
3	&  0.58\ud{0.06}{0.10}		& 83\ud{19}{13}	& 44\ppm17 		& 1.2\ppm0.6	   & 3\ud{5}{2}\tee{-5}& 18\%	& 19	&  0.98/9  (0.45)	\\
4	&  0.66\ud{0.04}{0.09}		& 67\ud{11}{11}	& 80\ud{44}{33}		& 1.9\ppm0.5	   & 1\ud{1}{0.5}\tee{-5}&  24\%& 23	&  1.3/9  (0.25)	\\
\enddata 
\tablecomments{
$^a$ X-ray fluxes are corrected for galactic absorption, in units of
\ee{-13} \cgsflux\ (0.5-10 keV).  $N_{pl}$ is the power-law component normalization in  phot
\perval{keV}{-1} \perval{cm}{-2} \perval{s}{-1} at 1 keV.  $f_{pl}$ is the fraction of the the
Total Model Flux which is accounted for by the power-law component
(these are typically uncertain by 50-100\%).
Upper-limits are 90\% confidence,  uncertainties are 1$\sigma$. Values in
parenthesis are held fixed.  Assumed source distance D=5 kpc.}
\end{deluxetable}

\begin{deluxetable}{ccc}
\tablecaption{\label{tab:nh} \nh\ in the direction of \aql }
\tablewidth{18cm}
\tablehead{
\colhead{Ref. } &
\colhead{Method} &
\colhead{Value} 
}
\startdata
\acite{callanan99} 	& Optical Photometry  & 0.29\ppm0.06	\\
\acite{dickey90} (W3NH)	& Weighted average integrated 21 cm emission & 0.34\ppm0.01	\\
\acite{thorstensen78}	& Optical Photometry of nearby B star & 0.40 \\
Obs. 1+2$^a$ (0.3-9 keV) & X-ray spectral modelling	& 0.43\ppm0.03 \\
Obs. 3+4$^a$ (0.3-9 keV) & X-ray spectral modelling	& 0.61\ppm0.06 \\
\enddata 
\tablecomments{The value for observations 3+4, without accounting for
the apparent 0.45-0.6 keV deficit,  appears discrepant}
\end{deluxetable}

\begin{deluxetable}{cccccccc}
\scriptsize
\tablecaption{\label{tab:specalla} Spectral Parameters for Observations
1+2 and 3+4}
\tablehead{
\colhead{} & 
\colhead{} & 
\colhead{\kteffinfty} &  
\colhead{\rinfty}  & 
\colhead{} & 
\colhead{$N_{pl}$} & 
\colhead{$f_{pl}$} & 
\colhead{} \\
\colhead{Obs. \#} & 
\colhead{\nhtt} & 
\colhead{(eV)} &  
\colhead{(km/(D/ 5 kpc))}  & 
\colhead{$\alpha$} & 
\colhead{(phot/\perval{cm}{-2}\perval{s}{-1} \@ 1 keV)} & 
\colhead{(\%)} &
\colhead{\chisqrnu\ (prob)} 
}
\startdata
\multicolumn{8}{c}{\dotfill Observations 1+2; \nh\ varies\dotfill  }\\ 
1	&	n/a	 	 	& n/a		& n/a			& n/a	    	    &n/a	& n/a			& 5.0/22  (1\tee{-13})	\\
2	&	n/a			&	''	& ''			& n/a	    	    &n/a	& n/a			&	``		\\ 
\multicolumn{8}{c}{\dotfill Observations 1+2; \rinfty\ varies\dotfill  }\\ 
1	&	0.41\ppm0.03 	 	& 122\ud{11}{9}	& 18.0\ppm3.5		& 0.4\ppm0.9        &$<$4.5\tee{-5}    &  $<$11         & 0.83/22 (0.69)		\\
2	&	''			&	``	& 12.9\ppm2.6		& ''                &     ''           &  $<$21         &	``		\\ 
\multicolumn{8}{c}{\dotfill Observations 1+2; \kteffinfty\ varies\dotfill  }\\ 
1	&	0.43\ppm0.03 	 	& 125\ud{12}{9}	& 17.2\ud{2.6}{3.6}	& 0.9\ppm1.0	    &$<$9\tee{-5}	& $<$21		& 0.67/22 (0.87)		\\
2	&	``			& 108\ud{11}{7} & ''			& ''		    & ''		& $<$35		&	``		\\ 
\multicolumn{8}{c}{\dotfill Observations 1+2; \fxpl\ varies\dotfill  }\\ 
1	&	0.83\ppm0.04		& (115)	& \rinfty$<$7.1			& 4.5\ppm0.2	    &(2.1\ppm0.3)\tee{-3}& $>$97		& 1.2/22 (0.36)		\\ 
2	&	``			& ``		 & ''			& ''		    & (1.1\ppm0.2)\tee{-3}& $>$95		&	``		\\ 
\multicolumn{8}{c}{\dotfill Observations 3+4 \dotfill }\\ 
3+4	&	0.60\ppm0.05 	 	& 81\ppm12 	& 46\ud{23}{11} 	& 1.5\ppm0.5	    &(5\ppm2)\tee{-5}	& 21 		& 1.2/23 (0.25)	\\
3+4	&       (0.43)			& 113\ppm1	& (17.2)		& 0.8\ppm0.3	    &(1.7\ud{1.0}{0.7})\tee{-5} & 26    & 1.71/25 (0.015)\\  
3+4$^b$ & (0.43)	& 114\ppm1	& (17.2)
& 0.7\ppm0.4	    &(1.5\ud{1.1}{0.7})\tee{-5} & 26    & 0.76/23
(0.78)\\  \hline
\enddata 
\tablecomments{
Uncertainties are 1$\sigma$, upper-limits are 90\%
confidence. Assumed distance is 5 \kpc \cite{rutledge01b}. Notation
``n/a'' (not applicable) is used where spectral parameters were used
in a fit which is not statistically acceptable. Upper-limits on $f_{pl}$,  the fraction of the 0.5-10.0 unabsorbed flux due to the power-law component, 
were derived from the best-fit model for the normalization $N_{pl}$ held fixed its 90\% confidence limit.}
\end{deluxetable}

\begin{deluxetable}{cccccccc}
\scriptsize
\tablecaption{\label{tab:specallb} Joint Spectral Parameters
(Observations 1-4)}
\tablehead{
\colhead{} & 
\colhead{} & 
\colhead{\kteffinfty} &  
\colhead{\rinfty}  & 
\colhead{} & 
\colhead{$N_{pl}$} & 
\colhead{$f_{pl}$} & 
\colhead{} \\
\colhead{Obs. \#} & 
\colhead{\nhtt} & 
\colhead{(eV)} &  
\colhead{(km/(D/ 5 kpc))}  & 
\colhead{$\alpha$} & 
\colhead{(phot/\perval{cm}{-2}\perval{s}{-1} \@ 1 keV)} & 
\colhead{(\%)} &
\colhead{\chisqrnu\ (prob)} 
}
\startdata
\multicolumn{8}{c}{\dotfill Observations 1-4  (0.6-9.0 keV); \kteffinfty\ varies \dotfill } \\ 
1	& 0.42\ppm0.03  		& 121\ud{13}{6}	& 18.4\ud{0.4}{3.8}	& 1.1\ud{0.2}{0.6} & (1.7\ud{1.4}{0.9})\tee{-5}& 13        	& 1.0/41 (0.40)  \\
2	& ''				& 105\ud{9}{6}	& ''			& ''		   & ''			    & 23        	& ''			\\
3+4	& ''				& 110\ud{12}{5}	& ''			& ''		   & ''			    & 19        	& ''			\\ 
\multicolumn{8}{c}{\dotfill Observations 1-4  (0.6-9.0 keV); \nh\ varies \dotfill } \\ 
1	& n/a				& n/a		& n/a			& n/a		   & n/a		& n/a       	& 3.0/41 (2\tee{-10})   \\
2	& n/a				& ''		& ''			& ''		   & ``			& ``        	& ''			\\
3+4	& n/a   			& ''		& ''			& ''		   & ``			& ``        	& ''			\\  
\multicolumn{8}{c}{\dotfill Observations 1-4  (0.6-9.0 keV); \rinfty\ varies\dotfill } \\  
1	& 0.39\ud{5}{3}	  		& 122\ud{13}{17}	& 17.6\ud{4.3}{2.5}& 0.7\ud{0.5}{0.3} & (0.8\ud{1.4}{0.4})\tee{-6}&  13       	& 1.13/41 (0.26)  \\
2	& ''				& ``			& 12.5\ud{3.1}{1.8}	& ''		   & ''			& 23        	& ''			\\
3+4	& ''				& ``			& 14.3\ud{1.6}{2.1}	& ''		   & ''			& 18        	& ''			\\  
\multicolumn{8}{c}{\dotfill Observations 1-4  (0.6-9.0 keV); \fxpl varies\dotfill } \\  
1	& n/a				& n/a		& n/a			& n/a		   & n/a       		& n/a                & 1.60/41 (0.0085)   \\
2	& ''				& ''		& ''			& ''		   & n/a        	& n/a                & ''			\\
3+4	& ''				& ''		& ''			& ''		   & n/a        	& n/a                & ''			\\  
\multicolumn{8}{c}{\dotfill Observations 1-4  (0.6-9.0 keV);  $\alpha$ and \fxpl\ vary\dotfill } \\  
1	& 0.52\ppm0.04			& 88\ud{10}{7}	& 32.5\ud{4}{4}		& 3.2\ppm0.2	   & (4.5\ppm0.5)\tee{-4}&   54   		& 1.08/39 (0.34)   \\
2	& ''				& ''		& ''			& 1.0\ppm0.7	   & (9\ud{17}{6})\tee{-6}&  10    		& ''			\\
3+4	& ''				& ''		& ''			& 2.0\ppm0.3	   & (9.5\ppm3)\tee{-5}	&   30      		& ''			\\  

\multicolumn{8}{c}{\dotfill Observations 1-4  (0.6-9.0 keV);  \nh\ and $\alpha$ fixed, all other values vary\dotfill } \\  
1	& (0.34)			& 142\ud{4}{3}		&  11.9\ud{0.2}{0.3}	& (1.0)		   & $<$1.6\tee{-5}		&  $<$16   		& 0.73/41 (0.89)   \\
2	& ''				& 122\ud{4}{3}		& ''			& ``		   & $<$1.1\tee{-5}		&  $<$20    		& ''			\\
3+4	& ''				& 128\ppm4		& ''			& `` 		   & (2.0\ud{0.4}{0.1})\tee{-5}	&  28\ud{6}{3} 		& ''			\\  
\multicolumn{8}{c}{\dotfill Observations 1-4  (0.6-9.0 keV); \kteff\ varies, \nh\ and \rinfty\ fixed\dotfill } \\  
1	& (0.34)			& 138\ppm1	& (13.0)		& 0.6\ppm0.4	   & (0)			& (0) 		& 1.30/43 (0.11)   \\
2	& ''				& 120\ppm1	& ''			& ''		   & (0)			& (0)        	& ''			\\
3+4	& ''				& 124\ud{1}{2}	& ''			& ''		   & (1.1\ud{1.5}{0.3})\tee{-5}	& 29        	& ''			\\  
\multicolumn{8}{c}{\dotfill Observations 1-4  (0.3-9.0 keV); \kteff\ varies, \nh\ and \rinfty\ fixed\dotfill } \\  
1	& (0.34)			& n/a		& (13.0)		& n/a		   & (0)			& (0) 		& 2.63/51 (2\tee{-9})  \\
2	& ''				& n/a		& ''			& ''		   & (0)			& (0)        	& ''			\\
3+4	& ''				& n/a		& ''			& ''		   & n/a			& n/a        	& ''			\\  
\multicolumn{8}{c}{\dotfill Observations 1-4  (0.3-9.0 keV); \kteff\ varies, \nh\ and \rinfty\ float\dotfill } \\  
1	& n/a				& n/a		& n/a			& n/a		   & (0)			& (0) 		& 1.62/49 (0.0038)  \\
2	& ''				& n/a		& ''			& ''		   & (0)			& (0)        	& ''			\\
3+4	& ''				& n/a		& ''			& ''		   & n/a			& n/a        	& ''			\\  
\multicolumn{8}{c}\dotfill {$^a$ Observations 1-4  (0.3-0.45, 0.6-9.0 keV); \kteff\ varies, \nh\ and \rinfty\ float\dotfill } \\  
1	& 0.42\ud{0.02}{0.03}		&130\ud{3}{5}   & 15.9\ud{0.8}{2.9}	& 0.7\ppm0.2	   & (0)			& (0) 		& 1.26/48 (0.105)  \\
2	& ''				& 113\ud{3}{4}	& ''			& ''		   & (0)			& (0)        	& ''			\\
3+4	& ''				& 118\ud{9}{4}	& ''			& ''		   & (1.4\ud{0.9}{0.8})\tee{-5}	& 26        	& ''			\\  
\multicolumn{8}{c}{\dotfill Observations 1-4  (0.3-9.0 keV); \kteff\ varies, \nh\ and \rinfty\ float\dotfill } \\  
1	& n/a				& n/a		& n/a			& n/a		   & (0)			& (0) 		& 1.66/47 (0.0030)  \\
2	& n/a				& n/a		& ''			& ''		   & (0)			& (0)        	& ''			\\
3+4	& n/a				& n/a		& ''			& ''		   & n/a			& n/a        	& ''			\\  \hline
\enddata 
\tablecomments{$^a$ See Table~\ref{tab:lx} for luminosities derived
from these values.  $^b$. The 0.45-0.6 keV band is excluded from the
data fit with these models.
Uncertainties are 1$\sigma$, upper-limits are 90\%
confidence. Assumed distance is 5 \kpc \cite{rutledge01b}. Notation
``n/a'' (not applicable) is used where spectral parameters were used
in a fit which is not statistically acceptable. Upper-limits on $f_{pl}$,  the fraction of the 0.5-10.0 unabsorbed flux due to the power-law component, 
were derived from the best-fit model for the normalization $N_{pl}$ held fixed its 90\% confidence limit.}
\end{deluxetable}

\begin{deluxetable}{lcc}
\tablecaption{\label{tab:lx} X-ray Luminosities}
\tablewidth{6cm}
\tablehead{
\colhead{Obs. } & 
\colhead{Thermal $L_x$} &
\colhead{Total $L_x$} \\
}
\startdata
1	&       4.6     &	4.6  \\
2	&       2.5     &	2.5	\\
3+4	&       3.0     &	4.0	\\ 
\enddata 
\tablecomments{
Assumed source distance d=5 kpc.  Luminosities are corrected for
absorption, in units of \ee{33}
\cgslum, 0.5-10 keV.}
\end{deluxetable}

\begin{deluxetable}{lcc}
\tablecaption{\label{tab:counts} Observed vs. Best-Fit Model Counts in 0.45-0.6 keV Bin}
\tablewidth{12cm}
\tablehead{
\colhead{Obs.($i$)} & 
\colhead{Observed Counts ($x_i$)} &
\colhead{Model Counts ($\mu_i$)} 
}
\startdata
1&	23	&	26.0	\\
2&	23	&	19.5 \\
3&	13	&	21.4 \\
4&	10	&	26.7\\
\enddata 
\tablecomments{Best-Fit Model parameters used to derive these values are in
Table~\ref{tab:specallb}; the best-fit model for observations 1-4,
using 0.3-9.0 keV, ignoring 0.45-0.6 keV, permitting \kteffinfty\
to vary between observations, and \nh\ and \rinfty\ to be the same
between observations.  Best fit
\chisqrnu=1.26/48 dof, prob=0.105. }
\end{deluxetable}

\end{document}